\documentclass{aastex}
\usepackage{spr-astr-addons}
\usepackage{url}
\usepackage{amsfonts}
\usepackage{amssymb,epsfig}
\usepackage{latexsym}
\usepackage{amsmath}
\usepackage{graphicx}

\RequirePackage{color}

\begin{document}

\title{Statefinder diagnostic of logarithmic entropy corrected holographic dark energy with Granda-Oliveros IR cut-off}
\author{A. Khodam-Mohammadi $^{1}$,
~Antonio Pasqua $^2$, ~M. Malekjani $^1$, ~Iuliia Khomenko $^3$, ~M.
Monshizadeh $^4$}

 \affil{$^1$Physics Department,
Faculty of Science, Bu-Ali Sina University, Hamedan 65178, Iran
\\ $^2$Department of Physics, University of
Trieste, Via Valerio, 2 34127 Trieste, Italy \\ $^3$ Heat-and-Power
Engineering Department National Technical University of Ukraine
``Kyiv Politechnical Institute" Kiev, Ukraine \\ $^4$Physics
Department, Faculty of Science, Islamic Azad University, Hamedan
branch, Iran}

\altaffiltext{}{khodam@basu.ac.ir.}
\altaffiltext{}{toto.pasqua@gmail.com}
 \altaffiltext{}{malekjani@basu.ac.ir.}\altaffiltext{}{ju.khomenko@gmail.com}\altaffiltext{}{monshizadeh470@yahoo.com}

\begin{abstract}In this work, we have studied the logarithmic entropy
corrected holographic dark energy (LECHDE) model with
Granda-Oliveros (G-O) IR cutoff. The evolution of dark energy (DE)
density $\Omega'_D$, the deceleration parameter, $q$, and equation
of state parameter (EoS), $\omega_{\Lambda}$, are calculated. We
show that the phantom divide may be crossed by choosing proper model
parameters, even in absence of any interaction between dark energy
and dark matter. By studying the statefinder diagnostic and
$\omega_{\Lambda}-\omega_{\Lambda}^{\prime}$ analysis, the pair
parameters $\{r,s\}$ and
$(\omega_{\Lambda}-\omega_{\Lambda}^{\prime})$ is calculated for
flat GO-LECHDE universe. At present time, the pair $\{r,s\}$ can
mimic the $\Lambda$CDM scenario for a value of $\alpha/\beta\simeq
0.87$, which is lower than the corresponding one for observational
data ($\alpha/\beta=1.76$) and for Ricci scale ($\alpha/\beta=2$).
We find that at present, by taking the various values of
($\alpha/\beta$), the different points in $r-s$ and
$(\omega_{\Lambda}-\omega_{\Lambda}^{\prime})$ plans are given.
Moreover, in the limiting case for a flat dark dominated universe at
infinity ($t\rightarrow \infty$), we calculate $\{r,s\}$ at G-O
scale. For Ricci scale ($\alpha = 2$, $\beta = 1$) we obtain
$\{r=0,s=2/3\}$.

\end{abstract}
%

\section{Introduction}

It is widely accepted among cosmologists and astrophysicists that
our universe is experiencing an accelerated expansion. The evidences
of this accelerated expansion are given by numerous and
complementary cosmological observations, like the SNIa
\citep{perlmutter,astier}, the CMB anisotropy,
observed mainly by WMAP (Wilkinson Microwave Anisotropy Probe) \citep%
{bennett-09-2003,spergel-09-2003}, the Large Scale Structure (LSS) \citep%
{tegmark,abz1,abz2} and X-ray \citep{allen} experiments. \newline In
the framework of standard Friedmann-Lemaitre-Robertson-Walker (FLRW)
cosmology, a missing energy component with negative pressure (known
as Dark Energy (DE)) is the source of this expansion. Careful
analysis of cosmological observations, in particular of WMAP data
\citep{bennett-09-2003,spergel-09-2003,peiris} indicates that almost
70 percent of the total energy of the universe is occupied by DE,
whereas DM occupies almost the rest (the barionic matter represents
only a few percent of the total energy density). The contribution of
the radiation is practically negligible.\newline The nature of DE is
still unknown and many candidates have been proposed in order to
describe it (see \citep{copeland-2006,Padmanabhan-07-2003,peebles}
and references therein for good reviews).\newline The
time-independent cosmological constant $\Lambda $ with equation of
state (EoS) parameter $\omega =-1$ is the earliest and simplest DE
candidate. However, cosmologists know that $\Lambda $ suffers from
two main difficulties: the fine-tuning and the cosmic coincidence
problems \citep{copeland-2006}. The former asks why the vacuum
energy density is so small (about $10^{-123}$ times smaller than
what we observe) \citep{weinberg} and the latter says why vacuum
energy and DM are nearly equal today (which represents an incredible
coincidence if no internal connections between them are
present).\newline Alternative candidates for DE problem are the
dynamical DE scenarios with no longer constant but time-varying
$\omega $. It has been shown by observational data analysis of
SNe-Ia that the time-varying DE models give a better fit compared
with a cosmological constant. A good review about the problem of DE,
including a survey of some theoretical models, can be found in
\citep{miao}.\newline An important advance in the study of black
hole theory and string theory is the suggestion of the so called
holographic principle: according to it, the number of degrees of
freedom of a physical system should be finite, it should scale with
its bounding area rather than with its volume \citep{thooft} and it
should be constrained by an infrared cut-off \citep{cohen}. The
Holographic DE (HDE), based on the holographic principle proposed by \citep%
{fischler}, is one of the most interesting DE candidates and it has
been widely studied in literature
\citep{enqvist-02-2005,shen,zhangX-08-2005,zhangX-11-2006%
,sheykhi-03-01-2010,huang-08-2004,hsu,guberina-05-2005,guberina-05-2006,gong-09-2004,%
elizalde-05-2005,jamil-01-2010,jamil1,jamil2,jamil3,%
jamil4,jamil5,setare-11-2006,setare-01-2007,setare-05-2007%
,setare-01-05-2007,setare-09-2007,setare-10-2007,setare-11-2007,%
setare-08-2008,setare-02-2010,2011setare,2011khodam,sheykhi-11-2009}.
The HDE model have also been constrained and tested by various astronomical observations \citep%
{enqvist-02-2005,shen,zhangX-08-2005,zhangX-07-2007,feng-02-2005,kao,micheletti,wangY,zhangX-05-2009}
as well as by the anthropic principle \citep{huang-03-2005}.\newline
Applying the holographic principle to cosmology, the upper bound of
the entropy contained in the universe can be obtained
\citep{fischler}. Following this line, \citep{li-12-2004} suggested
the following constraint on the energy density:
\begin{equation}
\rho _{\Lambda }\leq 3c^{2}M_{p}^{2}L^{-2},
\end{equation}
where $c$ is a numerical constant, $L$ indicates the IR cut-off radius, $%
M_{p}=(8\pi G)^{-1/2}\simeq 10^{18}$GeV is the reduced Planck mass ($G$ is
the gravitational constant) and the equality sign holds only when the
holographic bound is saturated. Obviously, in the derivation of HDE, the
black hole entropy (denoted with $S_{BH}$) plays an important role. As it is
well known, $S_{BH}=A/(4G)$, where $A\approx L^{2}$ is the area of the
horizon. However, this entropy-area relation can be modified as \citep%
{banerjee-04-2008,banerjee-06-2008,banerjee-05-2009}:
\begin{equation}
S_{BH}=\frac{A}{4G}+\tilde{\alpha}\log \left( \frac{A}{4G}\right) +\tilde{%
\beta},  \label{2}
\end{equation}%
where $\tilde{\alpha}$ and $\tilde{\beta}$ are dimensionless
constants. These corrections can appear in the black hole entropy in
Loop Quantum Gravity (LQG). They can also be due to quantum
fluctuation, thermal equilibrium fluctuation or mass and charge
fluctuations. The quantum corrections provided to the entropy-area
relationship leads to curvature correction in the Einstein-Hilbert
action and viceversa \citep{cai-08-2009,nojiri-2001,zhu}. Using the
corrected entropy-area relation given in Eq. (\ref{2}), the energy
density $\rho _{\Lambda }$ of the logarithmic entropy-corrected HDE
(LECHDE) can be written as \citep{wei-10-2009}:
\begin{equation}
\rho _{\Lambda }=3\alpha M_{p}^{2}L^{-2}+\gamma _{1}L^{-4}\log \left(
M_{p}^{2}L^{2}\right) +\gamma _{2}L^{-4},  \label{3}
\end{equation}%
where $\gamma _{1}$ and $\gamma _{2}$ are two dimensionless constants. In the
limiting case of $\gamma _{1}=\gamma _{2}=0$, Eq. (\ref{3}) yields the
well-known HDE density.\newline
The second and the third terms in Eq. (\ref{3}) are due to entropy
corrections: since they can be comparable to the first term only when $L$ is
very small, the corrections they produce make sense only at the early
evolutionary stage of the universe. When the universe becomes large, Eq. (%
\ref{3}) reduce to the ordinary HDE.\newline

It is worthwhile to mention that the IR cut-off $L$ plays an
important role in HDE model. By assuming particle horizon as IR
cut-off, the accelerated expansion can not be achieved \citep{hsu2},
while for Hubble scale, event horizon, apparent horizon and Ricci
scale, this fact may be achieved \citep{sheykhi-03-01-2010,pavon2,odintsov,pavon1,zimdahl}.\\
 Recently, Granda and Oliveros (G-O), proposed a new IR cut-off for HDE model, namely `new
holographic DE', which includes a term proportional to
$\overset{.}{H}\ \ $ and one proportional to $H^2$
\citep{grandaoliveros,granda2}. Despite of the HDE based on the
event horizon, this model depends on local
quantities, avoiding in this way the causality problem.\\
The investigation of cosmological quantities such as the EoS
parameter $\omega _{\Lambda }$, deceleration parameter $q$ and
statefinder diagnosis have attracted a great deal of attention in
new cosmology. Since the various DE models give $H>0$ and $q<0$ at
the present time, the Hubble and deceleration parameters can not
discriminate various DE models. A higher order of time derivative of
scale factor is then required. Sahni et al. \citep{sahni} and Alam
et al. \citep{alam}, using the third time derivative of scale factor
$a\left( t  \right)$, introduced the statefinder pair \{r,s\} in
order to remove the degeneracy of $H$ and $q$ at the present time.
The statefinder pair is given by:
\begin{eqnarray}
r&=&\frac{\overset{...}{a}}{aH^{3}}, \label{3s1}\\
s&=&\frac{r-1}{3(q-1/2)}.  \label{3s}
\end{eqnarray}
Many authors have been studied the properties of various DE models
from the viewpoint of statefinder diagnostic
\citep{state1,state2,state3,state4,state5,state6}.

This paper is organized as follows. In Section 2, we describe the
physical contest we are working in and we derive the EoS parameter
$\omega _{\Lambda } $, the deceleration parameter $q$ and $\Omega
_{\Lambda }^{\prime }$ for GO-LECHDE model. In Section 3, the
statefinder diagnosis and $\omega-\omega^{}\prime$ analysis of this
model are investigated. We finished our work with some concluding
remarks.

\section{cosmological properties}
The energy density of GO-LECHDE in Planck mass unit (i.e. $M_P=1$)
is given by
\begin{eqnarray}
\rho _{\Lambda } =\frac{3}{L_{GO}^{2}}\left[ 1+\frac{1}{3}%
L_{GO}^{-2}\left( 2\gamma _{1}\log L_{GO}+\gamma _{2}\right) \right] =\frac{3%
}{L_{GO}^{2}}\Gamma
\end{eqnarray}%
where we defined $\Gamma =1+\frac{1}{3}L_{GO}^{-2}\left( 2\gamma
_{1}\log L_{GO}+\gamma _{2}\right) $ for simplicity.
 The Granda-Oliveros IR cutoff given by \citep{grandaoliveros,khodam}:
\begin{equation}
L_{GO}=\left( \alpha H^{2}+\beta \dot{H}\right) ^{-1/2},  \label{4}
\end{equation}%
where $\alpha $ and $\beta $ are two constant.\\ The line element of
FLRW universe is given by:
\begin{equation}
ds^{2}=-dt^{2}+a^{2}\left( t\right) \left( \frac{dr^{2}}{1-kr^{2}}%
+r^{2}\left( d\theta ^{2}+\sin ^{2}\theta d\varphi ^{2}\right) \right) ,
\label{7}
\end{equation}%
where $t$ is the cosmic time, $a\left( t\right) $ is a dimensionless scale
factor (which is function of the cosmic time $t$), $r$ is referred to the
radial component, $k$ is the curvature parameter which can assume the values
$-1,\,0$ and $+1$ which yield, respectively, a closed, a flat or an open
FLRW universe and $\left( \theta ,\varphi \right) $ are the angular
coordinates.\\
The Friedmann equation for non-flat universe dominated by DE
and DM has the form:
\begin{equation}
H^{2}+\frac{k}{a^{2}}=\frac{1}{3}\left( \rho _{\Lambda }+\rho _{m}\right) ,
\label{8}
\end{equation}%
where $\rho _{\Lambda }$ and $\rho _{m}$ are, respectively, the energy
densities of DE and DM.\newline
We also define the fractional energy densities for DM, curvature and DE,
respectively, as:
\begin{eqnarray}
\Omega _{m} &=&\frac{\rho _{m}}{\rho _{cr}}=\frac{\rho _{m}}{3H^{2}},
\label{9} \\
\Omega _{k} &=&\frac{\rho _{k}}{\rho _{cr}}=\frac{k}{H^{2}a^{2}},  \label{10}
\\
\Omega _{\Lambda } &=&\frac{\rho _{\Lambda }}{\rho _{cr}}=\frac{\rho
_{\Lambda }}{3H^{2}}  \notag \\
&=&L_{GO}^{-2}H^{-2}\Gamma ,  \label{11}
\end{eqnarray}%
where $\rho _{cr}=3H^{2}$ represents the critical energy density.
Recent observations reveal that $\Omega _{k}\cong 0.02$
\citep{sperge}, which support a closed universe with a small
positive curvature.\newline
Using the Friedmann equation given in Eq. (\ref{8}), Eqs. (\ref{9}), (\ref%
{10}) and (\ref{11}) yield:
\begin{equation}
1+\Omega _{k}=\Omega _{m}+\Omega _{\Lambda }.  \label{12}
\end{equation}%
In order to preserve the Bianchi identity or the local energy-momentum
conservation law, i.e. $\nabla _{\mu }T^{\mu \nu }=0$, the total energy
density $\rho _{tot}=\rho _{\Lambda }+\rho _{m}$ must satisfy the following
relation:
\begin{equation}
\dot{\rho}_{tot}+3H\left( 1+\omega _{tot}\right) \rho _{tot}=0,  \label{13}
\end{equation}%
where $\omega _{tot}\equiv p_{tot}/\rho _{tot}$ represents the total
EoS parameter. In an non-interacting scenario of DE-DM, the energy
densities of DE and DM $\rho _{\Lambda }$ and $\rho _{m}$ are
preserved separately and the equations of conservation assume the
following form:
\begin{eqnarray}
\dot{\rho}_{\Lambda } &+&3H\rho _{\Lambda }\left( 1+\omega _{\Lambda
}\right) =0,  \label{14} \\
\dot{\rho}_{m} &+&3H\rho _{m}=0.  \label{15}
\end{eqnarray}%
The derivative with respect to the cosmic time $t$ of $L_{GO}$ is given by:
\begin{equation}
\dot{L}_{GO}=-H^{3}L_{GO}^{3}\left( \alpha \frac{\dot{H}}{H^{2}}+\beta \frac{%
\ddot{H}}{2H^{3}}\right) .  \label{16}
\end{equation}%
Using Eq. (\ref{16}), the derivative with respect to the cosmic time $t$ of
the energy density $\rho _{\Lambda }$ given in Eq. (\ref{3}) can be written
as:
\begin{eqnarray}
\dot{\rho}_{\Lambda }&=&
6H^{3}\left( \alpha \frac{\dot{H}}{H^{2}}+\beta \frac{%
\ddot{H}}{2H^{3}}\right)\times \nonumber\\
&&\left\{1+\frac{1}{3}L_{GO}^{-2}\left[ \gamma _{1}\left( 4\log
L-1\right) +2\gamma _{2}\right] \right\}.  \label{17}
\end{eqnarray}%
Differentiating the Friedmann equation given in Eq. (\ref{8}) with
respect to the cosmic time $t$ and using Eqs. (\ref{11}),
(\ref{12}), (\ref{15}) and
(\ref{17}), we can write the term $\alpha \frac{\dot{H}}{H^{2}}+\beta \frac{%
\ddot{H}}{2H^{3}}$ as:
\begin{equation}
\alpha \frac{\dot{H}}{H^{2}}+\beta \frac{\ddot{H}}{2H^{3}}=\frac{1+\frac{%
\dot{H}}{H^{2}}+\left( \frac{u}{2}-1\right) \Omega _{\Lambda }}{\{1+\frac{1}{%
3}L_{GO}^{-2}\left[ \gamma _{1}\left( 4\log L_{GO}-1\right) +2\gamma _{2}%
\right] \}},  \label{18}
\end{equation}%
where $u=\rho _{m}/\rho _{\Lambda }=\Omega _{m}/\Omega _{\Lambda
}=(1+\Omega _{k})/\Omega _{\Lambda }-1$ is the ratio of energy
densities of DM and DE. Using the expression of $L_{GO}$ given in
Eq. (\ref{4}) and the energy density of DE given in Eq. (\ref{7}),
we obtain that the term $\frac{\dot{H}}{H^2}$ can be written as:
\begin{equation}
\frac{\dot{H}}{H^{2}}=\frac{1}{\beta }\left( \frac{\Omega _{\Lambda }}{%
\Gamma }-\alpha \right) .  \label{19}
\end{equation}%
Therefore, Eq. (\ref{17}) yields:
\begin{equation}
\dot{\rho}_{\Lambda }=\frac{6H^{3}\Omega _{\Lambda }}{\beta }\left( \frac{1}{%
\Gamma }-\frac{\alpha -\beta }{\Omega _{\Lambda }}+\frac{\beta \left(
u-2\right) }{2}\right) ,  \label{20}
\end{equation}%
\newline
Differentiating the expression of $\Omega _{\Lambda }$ given in Eq. (\ref{11}) with respect to the cosmic time $t$ and
using the relation $\dot{\Omega}_{\Lambda }=H\Omega _{\Lambda }^{\prime }$,
we obtain the evolution of the energy density parameter as follow:
\begin{equation}
\Omega _{\Lambda }^{\prime }=\frac{2\Omega _{\Lambda }}{\beta }\left( \frac{1%
}{\Gamma }-\frac{\alpha -\beta }{\Omega _{\Lambda }}+\frac{\beta u}{2}%
\right) .  \label{21}
\end{equation}%
The dot and the prime denote, respectively, the derivative with respect to
the cosmic time $t$ and the derivative with respect to $x=\ln a$.\newline
Finally, using Eqs. (\ref{11}), (\ref{14}) and (\ref{20}), the EoS parameter
$\omega _{\Lambda }$ and the deceleration parameter (defined as $q=-1-\frac{\dot{H}}{%
H^{2}}$) as functions of $\Omega _{\Lambda }$ and $\Gamma$ are given,
respectively, by:
\begin{equation}
\omega _{\Lambda }=-\frac{2}{3 \Omega _{\Lambda }}\left[
1-\frac{\alpha}{\beta}+\frac{\Omega _{\Lambda }}{\beta \Gamma }
\right]-\frac{1+u}{3} , \label{22}
\end{equation}%
\begin{equation}
q=\left( \frac{\alpha}{\beta}-1 -\frac{\Omega _{\Lambda }}{\beta\Gamma }%
\right).  \label{23}
\end{equation}%
We can easily observe that the EoS parameter $\omega_{\Lambda}$ and the
 deceleration parameter $q$ given, respectively, in Eqs. (\ref{22}) and
 (\ref{23}) are related each other by the following relation:
\begin{equation}
\omega _{\Lambda }=\frac{2}{3\Omega _{\Lambda }}q-\frac{1+u}{3}.  \label{23a}
\end{equation}%
Moreover, using Eqs. (\ref{11}) and (\ref{23}), we can derive that:
\begin{equation}
L_{GO}^{-2}H_{GO}^{-2}=\frac{\Omega _{\Lambda }}{\Gamma }=\alpha -\beta
-\beta q=\alpha -\beta \left( 1+q\right) .  \label{24}
\end{equation}%
From Eqs. (\ref{14}) and (\ref{15}), the evolution of $u$ is
governed by:
\begin{equation}
u^{\prime }=3u\omega _{\Lambda }.  \label{28}
\end{equation}%
At Ricci scale, i.e. when $\alpha =2$ and $\beta =1$, Eqs. (\ref{22}) and (%
\ref{23}) reduce, respectively, to:
\begin{equation}
\omega _{\Lambda }=-\frac{2}{3\Omega _{\Lambda }}\left( \frac{\Omega
_{\Lambda }}{\Gamma }-1\right)-\frac{1+u}{3} ,  \label{25}
\end{equation}%
\begin{equation}
q=1-\frac{\Omega _{\Lambda }}{\Gamma },  \label{26}
\end{equation}%
and the evolution of the energy density parameter given in Eq. (\ref{21}) reduces to:
\begin{equation}
\Omega _{\Lambda }^{\prime }=\left[ 2\left( \frac{\Omega _{\Lambda }}{\Gamma}-1\right) \right] +u\Omega _{\Lambda }=-\Omega _{\Lambda }(1+3\omega
_{\Lambda }).  \label{27}
\end{equation}%
By choosing the proper model parameters, it can be easily shown that
the equation of state parameter $\omega_{\Lambda }$ given in Eqs.
(\ref{22}) and (\ref{25}), may cross the phantom divide. Moreover,
from Eqs. (\ref{23}) and (\ref{26}), we can see that the transition
between deceleration to acceleration
phase can be happened for various model parameters.\\
In a flat dark dominated universe, i.e. when $\gamma _{1}=\gamma _{2}=0$ or at infinity ($t\rightarrow \infty$),
 $\Omega _{\Lambda }=1$, $\Omega _{k}=0$ and $u=0$, we find that the
Hubble parameter $H$ reduces to:
\begin{equation}
H=\frac{\beta }{\alpha -1}\left( \frac{1}{t}\right) .  \label{29}
\end{equation}%
Moreover, the EoS parameter $\omega _{\Lambda }$ and the deceleration
parameter $q$ given in Eqs. (\ref{22}) and (\ref{23}) reduce, respectively,
to:
\begin{eqnarray}
\omega _{\Lambda }^{\infty}&=&-\frac{2}{3}\left( \frac{1-\alpha
}{\beta }\right) -1,
\label{30}\\
q^{\infty}&=&\frac{\alpha -1}{\beta }-1.  \label{31}
\end{eqnarray}%
Also in this case the phantom wall can be achieved for $\alpha \leq
1,~\beta>0$. In Ricci scale in this limit, Eqs. (\ref{30}),
(\ref{31}) reduce to
\begin{equation}
\omega_{\Lambda}^{R,\infty}=\frac{-1}{3},~~~q^{R,\infty}=0,
\end{equation}
which corresponds to an expanding universe without any acceleration.
\section{Statefinder diagnostic}

We now want to derive the statefinder parameters $\{r,s\}$ for
GO-LECHDE model in the
flat universe.\\
The Friedmann equation given in Eq. (\ref{8}) yields, after some calculations:
\begin{equation}
\frac{\dot{H}}{H^{2}}=-\frac{3}{2}\left( 1+\omega _{\Lambda }\Omega
_{\Lambda }\right) .  \label{s1}
\end{equation}%
Taking the time derivation of Eq. (\ref{s1}) and using Eq.
(\ref{21}), we obtain:
\begin{equation}
\frac{\ddot{H}}{H^{3}}=\frac{9}{2}\left[ 1+\omega _{\Lambda }^{2}\Omega
_{\Lambda }(1+\Omega )+\frac{7}{3}\omega _{\Lambda }\Omega _{\Lambda }-\frac{%
1}{3}\omega _{\Lambda }^{\prime }\Omega _{\Lambda }\right] .  \label{s2}
\end{equation}%
Using the definition of $H$ (i.e. $H=\dot{a}/a$), the statefinder parameter $r$ given in Eq. (\ref{3s1}) can be written as:
\begin{equation}
r=1+3\frac{\dot{H}}{H^{2}}+\frac{\ddot{H}}{H^{3}}.  \label{s3}
\end{equation}%
Substituting Eqs (\ref{19}), (\ref{23}) and (\ref{s2}) in Eqs.
(\ref{s3}) and (\ref{3s}), pair parameters $\{r,s\}$ can be written:
\begin{eqnarray}
r &=&1+6\omega _{\Lambda }\Omega _{\Lambda }+\frac{9}{2}\omega _{\Lambda
}^{2}\Omega _{\Lambda }(1+\Omega _{\Lambda })-\frac{3}{2}\omega _{\Lambda
}^{\prime }\Omega _{\Lambda },  \label{s4} \\
s &=&\beta \Gamma \Omega _{\Lambda }\left[ \frac{4\omega _{\Lambda }+3\omega
_{\Lambda }^{2}(1+\Omega _{\Lambda })-\omega _{\Lambda }^{\prime }}{\Gamma
(2\alpha -3\beta )-2\Omega _{\Lambda }}\right] \label{s4-1}.
\end{eqnarray}%
At early time, when $\omega_{\Lambda}\rightarrow 0$, the pair
relations (\ref{s4}) show that that statefinder parameters tends to
$\{r=1,s=0\}$, which coincides with the location of the $\Lambda$CDM
fixed point in $r-s$ plane.

Using Eq. (\ref{22}), the evolution of EoS parameter $\omega_{\Lambda}$  can be written as:
\begin{eqnarray}
\omega _{\Lambda }^{\prime }&=&\frac{2\Omega _{\Lambda }^{\prime
}}{3\beta
\Omega _{\Lambda }^{2}}\left( \frac{3}{2}\beta -\alpha \right)\notag \\&& +\frac{4}{%
3\beta \Gamma ^{2}}\left(\frac{L_{GO}^{\prime
}}{L_{GO}}\right)\left( 1+\frac{2\gamma _{1}}{3L_{GO}^{2}}-\Gamma
\right),\label{s5}
\end{eqnarray}%
where from Eqs. (\ref{11}) and (\ref{16}), the term
$\left(\frac{L_{GO}^{\prime }}{L_{GO}}\right)$ can be calculated as:
\begin{eqnarray}
\frac{L_{GO}^{\prime }}{L_{GO}} &=&-\frac{\Gamma }{\Omega _{\Lambda }}\left(
\alpha \frac{\dot{H}}{H^{2}}+\beta \frac{\ddot{H}}{2H^{3}}\right)  \label{s6}
\\
&=&\frac{3\Gamma }{2}\left\{ \frac{1+\omega _{\Lambda }}{1+\frac{1}{3}%
L_{GO}^{-2}\left[ \gamma _{1}\left( 4\log L_{GO}-1\right) +2\gamma _{2}%
\right] }\right\} .  \notag
\end{eqnarray}%
At present epoch of the Universe ($\Omega _{\Lambda }\approx 0.72$,
$u\approx 0.4)$, the EoS parameter $\omega _{\Lambda }$ given in Eq.
(\ref{23a}) reduces to:
\begin{equation}
\omega _{\Lambda }\approx 0.93q-0.47.
\end{equation}%
Then, the universe exists in accelerating phase (i.e $q<0$) if $\omega
_{\Lambda }<-0.47$ and the phantom divide $\omega _{\Lambda }=-1$, may be
crossed provided that $q\lesssim -0.5$. This condition implies $\frac{\dot{H}}{%
H^{2}}\gtrsim -0.58$ and, from Eq. (\ref{24}), we derive:
\begin{equation}
L_{GO-0}^{-2}H_{0}^{-2}\lesssim \alpha -0.42\beta,
\end{equation}
\begin{equation}
\Omega_{0 \Lambda }(\beta\Gamma _{0})^{-1}\gtrsim
\frac{\alpha}{\beta} -0.42 .
\end{equation}%
By inserting the above quantities in Eqs. (\ref{21}) and (\ref{s5}), we have $%
\omega _{\Lambda }^{\prime }\gtrsim -1.86\left( \alpha /\beta
-3/2\right) $, which gives:
\begin{eqnarray}
r_{0} &\approx &2\left(\frac{\alpha}{\beta}\right)-0.75, \\
s_{0} &\approx &-0.62\left(\frac{\alpha}{\beta}\right)+0.54.
\end{eqnarray}
Recently, Wang and Xu \citep{wangY} have constrained the new HDE
model in non-flat universe using observational data. The best fit
values of $( \alpha ,\beta ) $ with their confidence level they
found are $\alpha =0.8824_{-0.1163}^{+0.2180}( 1\sigma )
_{-0.1378}^{+0.2213}( 2\sigma ) $ and\\ $\beta
=0.5016_{-0.0871}^{+0.0973}( 1\sigma ) _{-0.1102}^{+0.1247}( 2\sigma
) $ . Using these values, the pair parameters $\{r,s\}$, at present
epoch, become $\{ r=2.77,s=-0.55\} $, which are far from $\Lambda
$CDM model values (i.e., $\left\{ r=1,s=0\right\} $). Moreover, it
shows that $s<0,$ which corresponds to a phantom-like DE. However,
in order to mimic these parameters to $\Lambda $CDM scenario at
present epoch, the ratio of $\alpha /\beta $ must be approximately
$0.87$,
which is lower than the value obtained with observational data.\\
At Ricci scale (i.e., when $\alpha /\beta=2$), at present time, pair parameters assume the values $%
\{r=3.25,s=-0.70\}$. It is worthwhile to mention that by increasing
the value of $\alpha /\beta$ from 0.87, the distance from $\Lambda
$CDM fixed point in $r-s$ diagram become longer.\\
In the limiting case of $t\rightarrow \infty $ or for ordinary new
HDE ($\gamma _{1}=\gamma _{2}=0,\Gamma =1$), in flat dark dominated
universe ($u=0,\Omega _{\Lambda }=1$), we find that:
\begin{eqnarray}
r &=&\frac{1}{\beta ^{2}}\left( \alpha -\beta -1\right) \left(
2\alpha
-\beta -4\right) \label{io3}, \\
s &=&\frac{2\left( 2\alpha ^{2}-3\beta \alpha +5\beta -6\alpha +4\right) }{%
3\beta \left( 2\alpha -3\beta -2\right) }\label{io4}.
\end{eqnarray}%
At Ricci scale ($\alpha =2,\beta =1$), Eqs. (\ref{io3}) and (\ref{io4}) reduce, respectively, to :%
\begin{eqnarray}
 r=0,~~s=\frac{2}{3}.
\end{eqnarray}
Moreover the $\omega-\omega^{\prime}$ analysis is another tool to
distinguish between the different models of DE \citep{wei2007}. In
this analysis the standard $\Lambda$CDM model corresponds to the
fixed point $(\omega_{\Lambda}=-1,\omega_{\Lambda}^{\prime}=0)$. At
present time, for $\alpha/\beta=0.87$ which corresponds
to$\Lambda$CDM fixed point in $r-s$ diagram,
$(\omega_{\Lambda}=-1,\omega_{\Lambda}^{\prime}=1.17)$. for the
observational quantities, ($\alpha/\beta=1.76$), we find:
$(\omega_{\Lambda}=-1,\omega_{\Lambda}^{\prime}=-0.48)$, and for
Ricci scale these are
$(\omega_{\Lambda}=-1,\omega_{\Lambda}^{\prime}=-0.93)$. Therefore
we see that $\omega_{\Lambda}^{\prime}$ become smaller for higher
value of $\alpha/\beta$ at present.

\section{Conclusion}

In this paper, we have extended the work made by Granda and Oliveros
\citep{grandaoliveros} to the logarithmic entropy corrected HDE
(LECHDE) model. This model has been arisen from the black hole
entropy which may lie in the entanglement of quantum field between
inside and outside of the horizon. We obtained the evolution of
energy density $\Omega _{\Lambda }^{\prime }$, the deceleration
parameter $q$ and EoS parameter $\omega_{\Lambda} $ of the new
LECHDE model for non-flat universe. We saw that, by choosing the
proper model parameters, the equation of state parameter $\omega
_{\Lambda }$ may cross the phantom divide and also the transition
between deceleration to acceleration phase could happen.

At last, we studied the GO-LECHDE model from the viewpoint of
statefinder diagnostic and
$\omega_{\Lambda}-\omega_{\Lambda}^{\prime}$ analysis, which is a
crucial tool for discriminating different DE models. Also, the
present value of $\{r, s\}$ can be viewed as a discriminator for
testing different DE models if it can be extracted from precise
observational data in a model-independent way. The studying at
present time, when $\omega_{\Lambda}$ remains around the phantom
wall, $\omega_{\Lambda}\approx -1$ and our universe evolves in
acceleration phase, pair values of $\{ r,s \}$ was calculated with
respect to model parameters $\alpha, \beta$. By using the
observational data which was obtained by Wang and Xu \citep{wangY},
where $\alpha/\beta=1.76$, we obtained $\{ r=2.77 ,s=-0.55\}$. For
Ricci scale, which has $\alpha/\beta=2$, the pair value assume the
values $\{ r=3.25,s=-0.7 \}$. Also, choosing $\alpha/\beta=0.87$, we
found $\{ r=1,s=0 \}$ which is corresponds to $\Lambda$CDM scenario.
We shaw that increasing value of $\alpha/ \beta$, conclude the
ascending distance from $\Lambda$CDM fixed point. In the limiting
case, at infinity, for flat dark dominated universe at Ricci scale,
we found $\{ r=0,s=2/3 \}$, which corresponds to an expanding
universe without any acceleration ($q=0$). In
$\omega_{\Lambda}-\omega_{\Lambda}^{\prime}$ analysis at present
time, we found that the higher value of $\alpha/\beta$ obtains the
smaller value of $\omega_{\Lambda}^{\prime}$.\\In this model the
statefinder pairs is determined by parameters
$\alpha,~\beta,~\gamma_1,~\gamma_2$. These parameters would be
obtained by confronting with cosmic observational data. Giving the
wide range of cosmological data available, in the future we expect
to further constrain our model parameter and test the viability of
our model.

\end{document}